\documentclass[11pt,a4paper]{book}

\usepackage{amssymb}
\usepackage{latexsym}
\usepackage{multirow,booktabs}
\usepackage{threeparttable}
\usepackage{graphicx}
\usepackage{psfrag,graphicx}
\usepackage{pstcol,pst-text}
\usepackage{amsmath}
\usepackage{natbib}
\usepackage{rotating,booktabs}
\usepackage{hyperref}

\usepackage{tkz-graph}
\usepackage{pstricks,pst-node,pst-tree}%

\usepackage{fancyhdr}

\def\mybibliography#1{{\begin{center} \bf References \end{center}}\list
 {}{\setlength{\leftmargin}{1em}\setlength{\labelsep}{0pt}
\itemindent=-\leftmargin}
 \def\newblock{\hskip .02em plus .20em minus -.07em}
 \sloppy\clubpenalty4000\widowpenalty4000
 \sfcode`\.=1000\relax}
\newbox\TempBox \newbox\TempBoxA
\def\uw#1{%
  \ifmmode\setbox\TempBox=\hbox{$#1$}\else\setbox\TempBox=\hbox{#1}\fi%
  \setbox\TempBoxA=\hbox to \wd\TempBox{\hss\char'176\hss}%
  \rlap{\copy\TempBox}\smash{\lower9pt\hbox{\copy\TempBoxA}}%
}
\newbox\TempBox \newbox\TempBoxA
\def\uwd#1{%
  \ifmmode\setbox\TempBox=\hbox{$#1$}\else\setbox\TempBox=\hbox{#1}\fi%
  \setbox\TempBoxA=\hbox to \wd\TempBox{\hss\char'176\hss}%
  \rlap{\copy\TempBox}\smash{\lower10pt\hbox{\copy\TempBoxA}}%
}
\def\mathunderaccent#1{\let\theaccent#1\mathpalette\putaccentunder}
\def\putaccentunder#1#2{\oalign{$#1#2$\crcr\hidewidth
\vbox to.2ex{\hbox{$#1\theaccent{}$}\vss}\hidewidth}}

\newcommand{\bee}{\begin{eqnarray*}}
\newcommand{\eee}{\end{eqnarray*}}
\newcommand{\be}{\begin{eqnarray}}
\newcommand{\ee}{\end{eqnarray}  }

\def\ni{\noindent}

\def\be{\begin{eqnarray*}}
\def\ee{\end{eqnarray*}}

\begin{document}

\thispagestyle{empty}

{\hbox{\footnotesize\rm Journal of Data Science {\footnotesize\bf
 Vol. 15}(2017), 347-356}\hfill}

\vspace{0.4pc}

\begin{center}
{\large\bf CONDITIONAL INDEPENDENCE TEST FOR CATEGORICAL DATA USING POISSON LOG-LINEAR MODEL }
\end{center}

\vspace{.2cm}
\begin{center}
\renewcommand{\thefootnote}{\fnsymbol{footnote}}

Michail\ Tsagris$^{1}$ \\
~\\

${\it} ^{1}$Department of Computer Science, University of Crete, Herakleion, Greece\\

\end{center}

{\small
\begin{quotation}
\ni {\it Abstract}:~~We demonstrate how to test for conditional independence of two variables with categorical data using Poisson log-linear models. The size of the conditioning set of variables can vary from 0 (simple independence) up to many variables. We also provide a function in R for performing the test. Instead of calculating all possible tables with \textit{for} loop we perform the test using the log-linear models and thus speeding up the process. Time comparison simulation studies are presented. 

\vspace{0.4cm}
\ni {\it Key words}:~~ Conditional independence, categorical data, Poisson log-linear models.\\
\end{quotation}
}

\vspace{0.4cm}
\noindent{\bf 1. Introduction}\\
When testing for (conditional independence) with categorical variables, the most famous test is the $G^2$ which is calibrated against a $\chi^2$ distribution with the appropriate degrees of freedom (Tsamardinos et al., 2006). But why is this issue so important? The reason is that when building Bayesian networks with categorical variables, conditional independence testing is a cornerstone (Neapolitan, 2004). 

All textbooks regarding categorical data analysis we came across, do mention the concept of independence and conditional independence. In addition, all of them have examples of testing whether two categorical variables are independent conditioning on another categorical variable. Agresti (2002) mentions an example where two variables compose belong in the conditioning set, but with not much details. We have tracked down two papers regarding Poisson log-linear models (Cheng et al., 2006; Cheng et al., 2007), but they do not convey the message we want to convey with this paper. 

The R package \href{https://cran.r-project.org/web/packages/pcalg/index.html}{pcalg} (Kalisch et al., 2012) offers the $G^2$ test of conditional independence for categorical data. But, the implementation of the test two drawbacks. It calculates the $G^2$ test but not the relevant degrees of freedom. Secondly it becomes slower as the sample size and or the conditioning set of variables increases. For these reasons, we decided to demonstrate how one can perform the $G^2$ and the $\chi^2$ test of conditional independence and in addition how both of them can be implemented in R using the \textit{MASS} library which is already installed in R.
 
The R package \textit{coin} (Zeileis et al., 2008) offers conditional independence tests, but the cardinality of the conditioning set is limited to 1, hence it is not considered any further in this work.

Section 2 describes the two tests of conditional independence and how they can be performed using log-linear models. Section 3 contains some simulation studies and finally Section 4 concludes the paper.  

\vspace{0.4cm}
\noindent{\bf 2. $G^2$ and $\chi^2$ tests of conditional independence}\\
\noindent{\bf 2.1 $G^2$ and $\chi^2$ tests of independence between two variables}\\

The $\chi^2$ and $G^2$ tests of independence between two categorical variables $X$ and $Y$ are defined as
\begin{eqnarray} \label{tests}
\chi^2 = \sum_{x, y}\frac{\left(N_{xy}-E_{xy}\right)^2}{E_{xy}} \ \ \ \text{and} \ \ \ 
G^2 = 2 \sum_{x, y}N_{xy}\log{\frac{N_{xy}}{E_{xy}}} 
\end{eqnarray}
respectively. Both of them follow an asymptotic $\chi^2$ distribution with $\left(\left|X\right|-1\right)\left(\left|Y\right|-1\right)$, 
where $\left| . \right|$ denotes the number of values of the variable. Since we can cross-tabulate the two variables, it is easier  to see the observed frequencies of all the pairs of the values of $X$ and $Y$, $N_{xy}$ and calculate their corresponding expected frequencies $E_{xy}$ under the assumption of independence as 
\begin{eqnarray*}
E_{xy} = \frac{N_{x+}N_{+y}}{N_{++}}, 
\end{eqnarray*} 
where $N_{x+}=\sum_{y}N_{xy}$, $N_{+y}=\sum_{x}N_{xy}$ and $N_{++}=\sum_{x, y}N_{xy}$. Note that $N_{++}=N$, the total sample size.    

An alternative way to calculate the expected frequencies and thus the values of $\chi^2$ and $G^2$ in (\ref{tests}) is via the Poisson log-linear models, hereafter denoted by PLL models for convenience purposes. If we fit a PPL model where the dependent variable is the obeserved frequencies $N$ and the variables $X$ and $Y$ play the role of the predictor variables we get
\begin{eqnarray} \label{pois}
\log{N_{ij}} = a + \beta X +\gamma Y + e
\end{eqnarray} 
The deviance of this model is equal to the value of the $\chi^2$ test statistic (Agresti, 2002). If we try to fit the model
\begin{eqnarray*}
\log{N_{ij}} = \beta_0 + \beta1 X +\beta_2 Y +\gamma X:Y + e,
\end{eqnarray*}
the deviance is zero, because this is the saturated model. It has as many parameters as observed frequencies in the contingency table (cross-tabulation of $X$ and $Y$). So, if the two variables are independent, the deviance of (\ref{pois}) should be small, meaning that the simple model is enough to predict the observed frequencies and so it fits adequately. Consequently, this leads to the conclusion that no interaction is significant and hence the two variables $X$ and $Y$ can be assumed independent. 

In order to calculate the value of the $G^2$ test statistic, we must use the predicted values of the simple model (\ref{pois}) 
$E_{xy}=\hat{N}=exp^{a + \beta X +\gamma Y }$ and plug them into the $G^2$ test formula in (\ref{tests}). If we plug them into the $\chi^2$ formula we get the deviance. So there is no reason to do it, since we already have the answer. 

\vspace{0.4cm}
\noindent{\bf 2.2 $G^2$ and $\chi^2$ tests of conditional independence between two variables conditioning on a third variable}\\

If we have a third variable $Z$ upon which we want to condition the independence of $X$ and $Y$ the two tests in Eq. (\ref{tests}) become (Tsamardinos and Borboudakis, 2010)
\begin{eqnarray} \label{citests}
\chi^2 = \sum_{x, y, {\bf z}}\frac{\left(N_{xy{\bf z}}-E_{xy{\bf z}}\right)^2}{E_{xy{\bf z}}} \ \ \ \text{and} \ \ \ 
G^2 = 2 \sum_{x, y {\bf z}}N_{xy{\bf z}}\log{\frac{N_{xy{\bf z}}}{E_{xy{\bf z}}}} 
\end{eqnarray}
respectively. Both of them follow an asymptotic $\chi^2$ distribution with $ \\ \left(\left|X\right|-1\right)\left(\left|Y\right|-1\right)\left|Z\right|$ degrees of freedom and the expected frequencies are calculated as
\begin{eqnarray*}
E_{xy{\bf z}} = \frac{N_{x+{\bf z}}N_{+y{\bf z}}}{N_{++{\bf z}}}, 
\end{eqnarray*} 
where $N_{x+{\bf z}}$, $N_{+y{\bf z}}$ and $N_{++{\bf z}}$ are the same as before and are calculated for every value of $Z$. The difference from before is that now instead of one contingency table we have $\left|Z\right|$ tables. We can have of course a 3-way table if we want. 

The PLL model to test the conditional independence of $X$ and $Y$ given $Z$ is 
\begin{eqnarray*}
\log{N_{ijk}} = \beta_0 + \beta_1 X +\beta_2 Y + \beta_3 Z + \gamma_1 X:Z + \gamma_2 Y:Z + e. 
\end{eqnarray*}
Again, the deviance of this model is equal to the value of the $\chi^2$ test statistic (Agresti 2002). We have included two interactions, one $X$ and $Z$ and one with $Y$ and $Z$. The interaction between $X$ and $Y$ is missing, as this would test for homogeneous association, but we are not interested in this one here. What is missing also is the interaction of $X$, $Y$ and $Z$ ($X:Y:Z$). If the deviance of the fitted model is small enough to say the model fits the data well, then the assumption of conditional independence is not rejected. In the same spirit, $\log{N} = a + \beta X +\gamma Y + \delta_1 Z:X + \delta_2 Y:X + e$ tests for the conditional independence of $Z$ and $Y$ given $X$ and $\log{N} = a + \beta X +\gamma Y + \delta_1 Z:Y + \delta_2 X:Y + e$ tests for the conditional independence of $Z$ and $X$ given $Y$. 

\vspace{0.4cm}
\noindent{\bf 2.3 $G^2$ and $\chi^2$ tests of conditional independence between two variables conditioning on a set of variables}\\

Moving on to higher order conditional independence tests, the procedure is the same, calculation of (\ref{citests}) where ${\bf Z}$ is a set of variables $(Z_1,\ldots,Z_k)$. This means that the degrees of freedom of the asymptotic $\chi^2$ distribution become
$\left(\left|X\right|-1\right)\left(\left|Y\right|-1\right)\prod_{i=1}^k\left|Z_i\right|$ (Tsamardinos et al., 2006).

It is clear now that calculation of the test statistics (\ref{tests}) becomes more difficult, since we have to make all combinations of the variables and produce the relevant contingency tables and so on. For this reason we will use again the PLL models. We could not find how to test for the higher order conditional independence in the textbooks and so we decided to show the way, or the general rule if your prefer. 

\vspace{0.4cm}
\noindent{\bf 2.3.1 Two conditioning variables}\\

Suppose ${\bf Z}=\left(Z_1, Z_2\right)$. The PLL we have to fit is
\begin{eqnarray*}
\log{N_{ijkl}} &=&  \beta_0 + \beta_1 X +\beta_2 Y + \beta_3 Z_1 + \beta_4 Z_2 \\ 
& & + \gamma_1 Z_1:Z_2+ \gamma_2 X:Z_1 + \gamma_3 X:Z_2+ \gamma_4 Y:Z_1 + \gamma_5 Y:Z_2   \\
& & + \delta_1 X:Z_1:Z2 + \delta_2 Y:Z_1:Z2 + e. 
\end{eqnarray*}

We have included all main effects (first row), all 2-way interactions between the variables except from the $X:Y$ (second row) and in the third row we have put the main 3-way interactions of interest. The 3-way interactions of $X$ and of $Y$ with the conditioning set of variables. 

\vspace{0.4cm}
\noindent{\bf 2.3.2 Three conditioning variables}\\

If now we have three variables in the conditioning set $${\bf Z}=\left(Z_1, Z_2, Z_3\right)$$ the PLL to be fitted is written as
\begin{eqnarray*}
\log{N_{ijklm}} &=&  \beta_0 + \beta_1 X +\beta_2 Y + \beta_3 Z_1 + \beta_4 Z_2 +\beta_5 Z_3 \\ 
& & +\gamma_1 Z_1:Z_2 + \gamma_2 Z_1:Z_3 +\gamma_3 Z_2:Z_3 + \gamma_4 Z_1:Z_2:Z_3  \\
& & + \gamma_5 X:Z_1 + \gamma_6 X:Z_2+ \gamma_7 X:Z_3 + \gamma_8 Y:Z_1 + \gamma_9 Y:Z_2 + \gamma_{10} Y:Z_3 \\
& & + \gamma_{11}X:Z_1:Z_2 + \gamma_{12}X:Z_1:Z_3 + \gamma_{13}X:Z_2:Z_3 \\
& & + \gamma_{14}Y:Z_1:Z_2 + \gamma_{15}Y:Z_1:Z_3 + \gamma_{16}Y:Z_2:Z_3 \\
& & + \delta_1 X:Z_1:Z_2:Z_3 + \delta_2 Y:Z_1:Z_2:Z_3 + e. 
\end{eqnarray*}
Note that we have included only up to 3-way interactions of $X$ and of $Y$ with the conditioning variables and not more than that. The final row is the row of interest. It is the 4-way interactions of $X$ and of $Y$ with all the conditioning variables.

\vspace{0.4cm}
\noindent{\bf 2.3.3 The general rule for $k$ conditioning variables}\\

The general model for $k < \infty$ conditioning variables (excluding the regression coefficients for convenience purposes) is given below
\begin{eqnarray*}
\log(N_{ij{\bf cs}}) &=& X+Y + \sum_{i=1}^kZ_i + \sum_{i\neq j}^kZ_i:Z_j + \sum_{i\neq j \neq l}^kZ_i:Z_j:Z_l + \ldots  \\
& & + \left(X + Y\right):\left(\sum_{i=1}^kZ_i + \sum_{i\neq j}^kZ_i:Z_j + 
\sum_{i\neq j \neq l}^kZ_i:Z_j:Z_l + \ldots \right) \\
& & + (X+Y):\left(Z_1:Z_2: \ldots :Z_k\right) +e.
\end{eqnarray*}

The first row contains the main effects and all the 2-way, 3-way, 4-way up to $k$-way interactions of the conditioning variables. 
The second row contains the 2-way, 3-way, up to the $k$-way interactions of the conditioning variables with the variables of interest $X$ and $Y$. Note that up to $k-1$ variables are included in the interactions. Finally, the last row of interest contains the $k+1$-way interactions of $X$ and $Y$ with the conditioning variables. 

We believe that our point is made clear now and the interesting reader can proceed to higher orders should he or she wish to. For this reason, we will proceed with the time comparisons between the functions available in the R package \textit{pcalg} and our function (which appears in the Appendix). 

\vspace{0.4cm}
\noindent{\bf 3. Time comparisons} \\

The R package \textit{pcalg} contains two functions, \textit{gSquareBin} when all the variables are binary and \textit{gSquareDis} for all other cases. In addition, the function \textit{disCItest} which is a wrapper of \textit{gSquareDis} will also be examined. The test based on the Poisson log-linear models will be denoted by PLL. We will focus only in the general case of arbitrary categories in each categorical variable and thus use the latter command. The time comparisons will take into account, three factors: the sample size, the number of times a test is implemented and the number of conditioning variables. For each of these cases and their combinations of course, $50$ repetitions will be made and the average times will be reported. The command in R used for this purpose is \textit{proc.time()}. We could also use \textit{system.time}, but we found the former one easier. 

The time required for $T=(500, 1000, 2000, 3000, 5000)$ tests was calculated. We used two categorical variables with 3 and 4 levels and conditioned on a) 1 variable with 2 levels (12 degrees of freedom), b) 2 variables with 2 and 4 levels (48 degrees of freedom) and c) 3 variables with 2, 4 and 4 levels (192 degrees of freedom). For all of these combinations, varying sample sizes $n=(3000, 5000, 10000)$ were chosen. All the results appear in Figure 1.

\begin{center} \vspace*{-0.2cm}
\begin{figure}[!ht]
\begin{tabular}{ccc}
\multicolumn{3}{c}{One conditioning variable}            \\
\includegraphics[scale=0.25,trim=0 20 20 20]{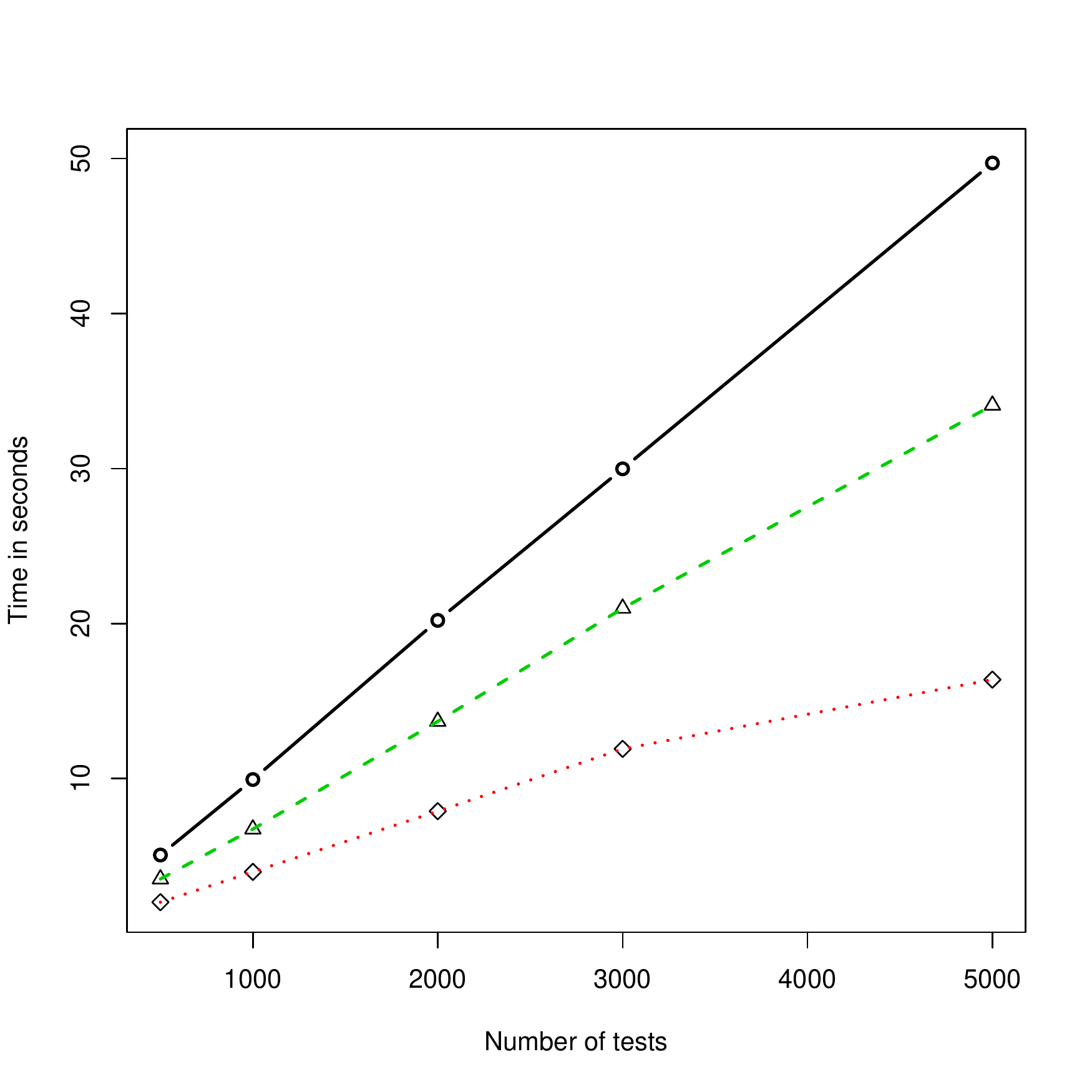}     &
\includegraphics[scale=0.25,trim=0 20 20 20]{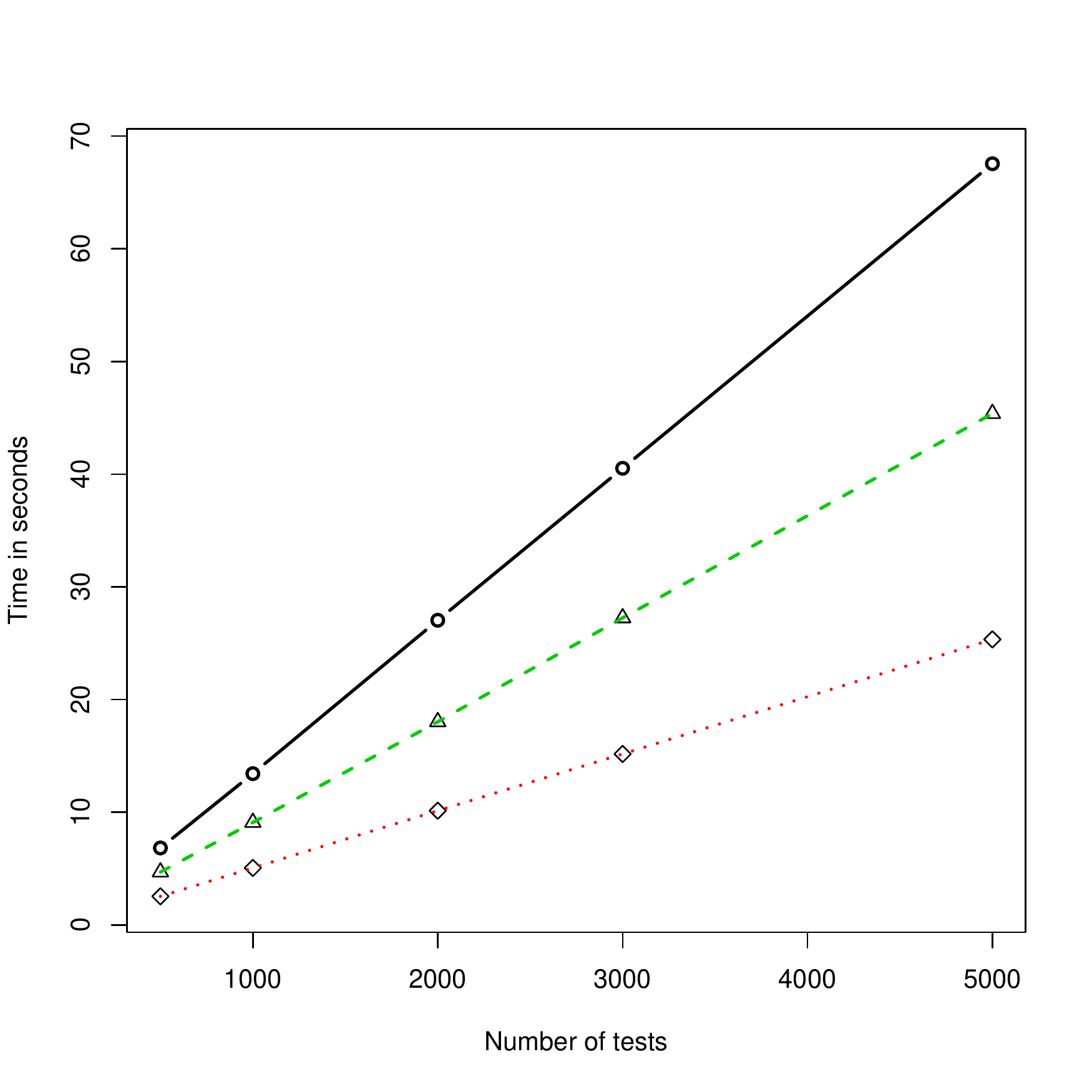}      &
\includegraphics[scale=0.25,trim=0 20 20 20]{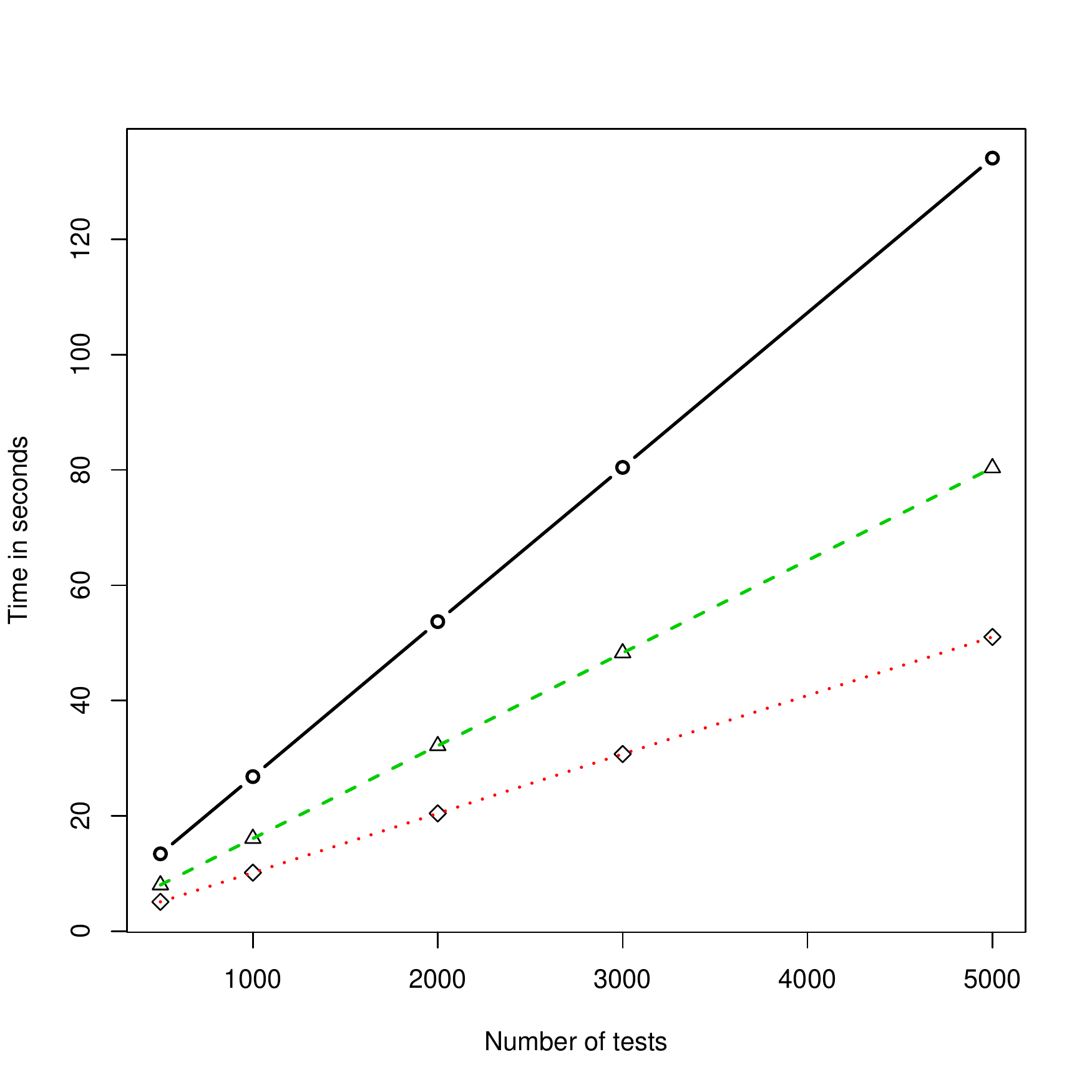}     \\
\footnotesize{$n=3000$}   &  \footnotesize{$n=5000$}    &  \footnotesize{$n=10000$}  \\  
\multicolumn{3}{c}{Two conditioning variables}           \\
\includegraphics[scale=0.25,trim=0 20 20 20]{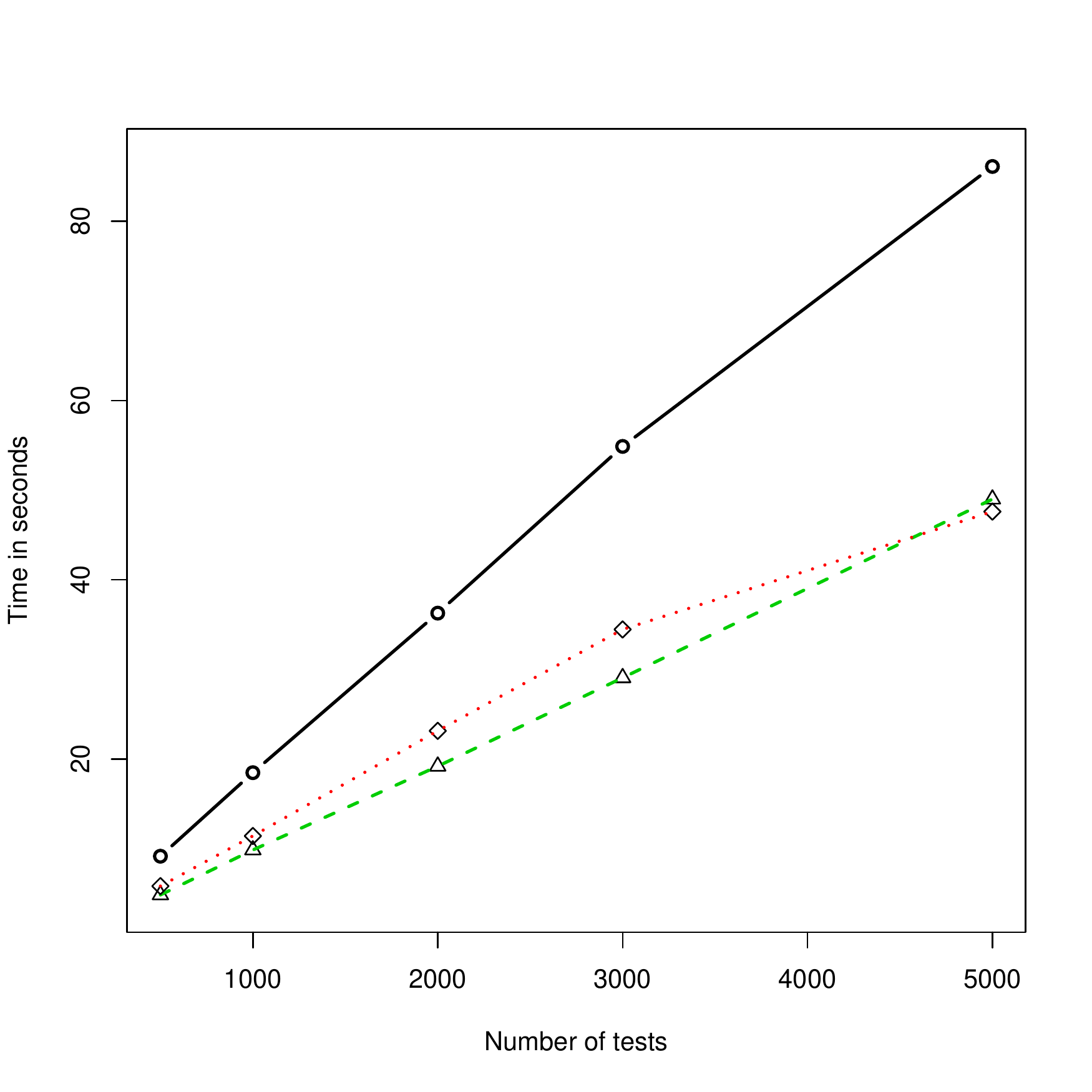}     &
\includegraphics[scale=0.25,trim=0 20 20 20]{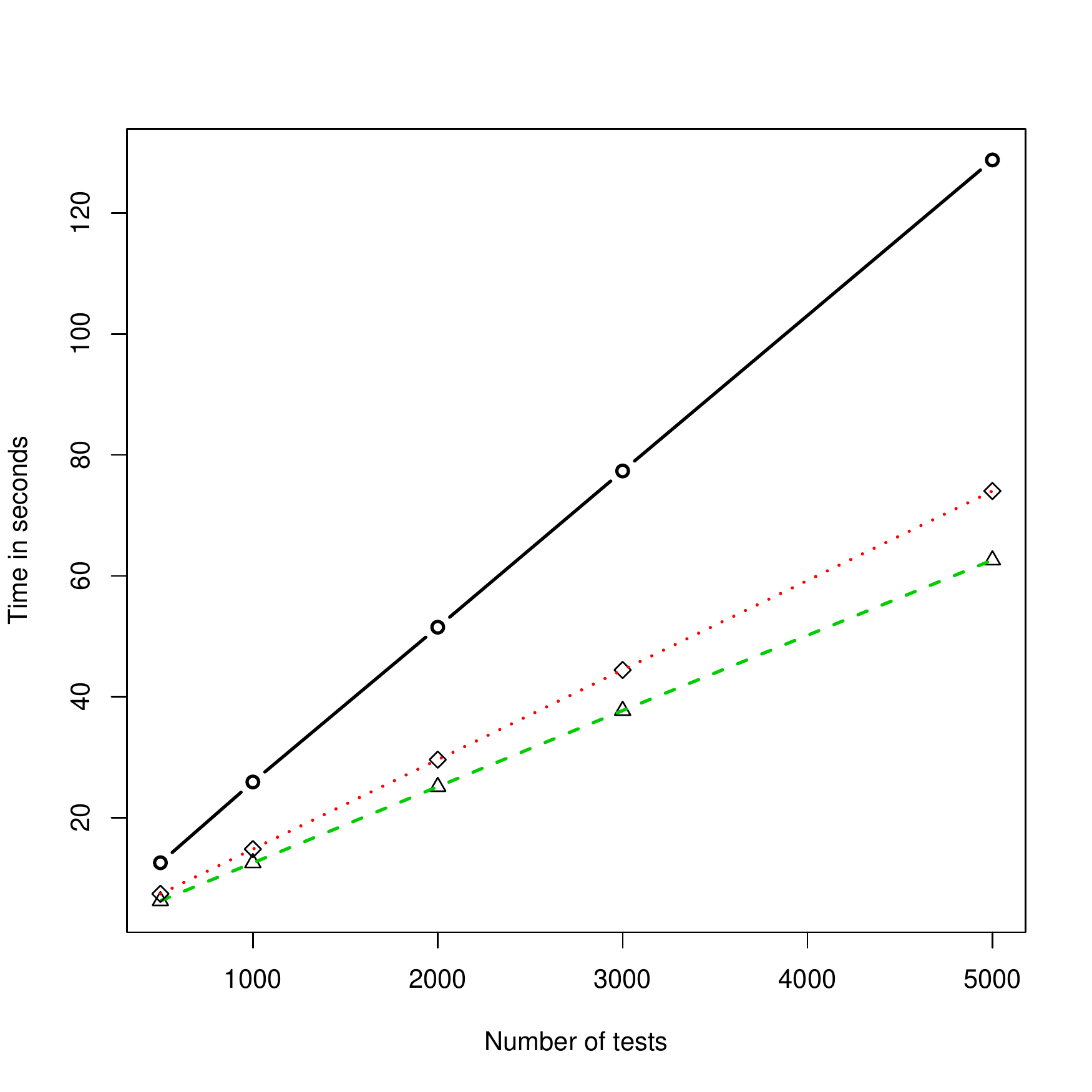}      &
\includegraphics[scale=0.25,trim=0 20 20 20]{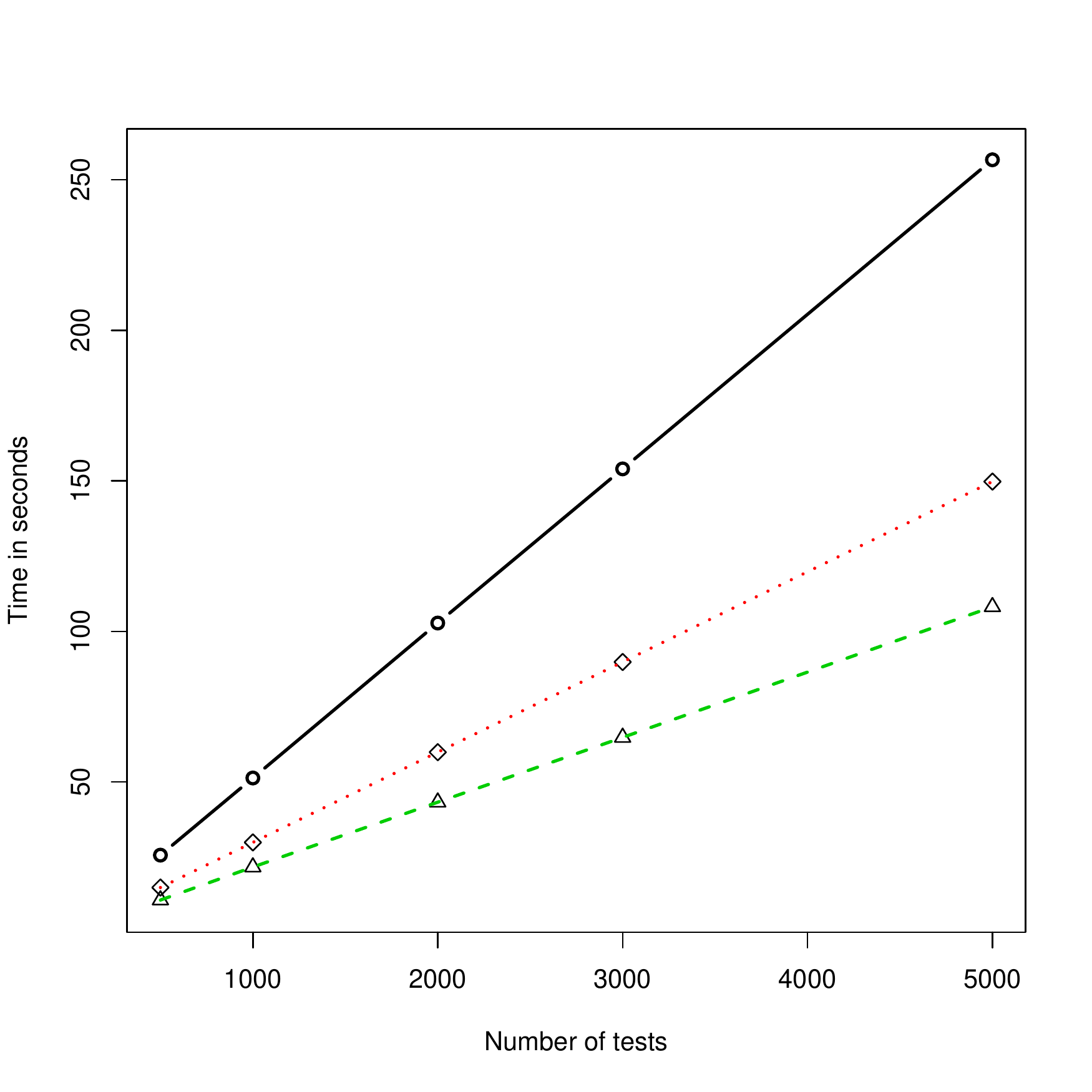}     \\
\footnotesize{$n=3000$}   &  \footnotesize{$n=5000$}  &  \footnotesize{$n=10000$}  \\  
\multicolumn{3}{c}{Three conditioning variables}         \\
\includegraphics[scale=0.25,trim=0 20 20 20]{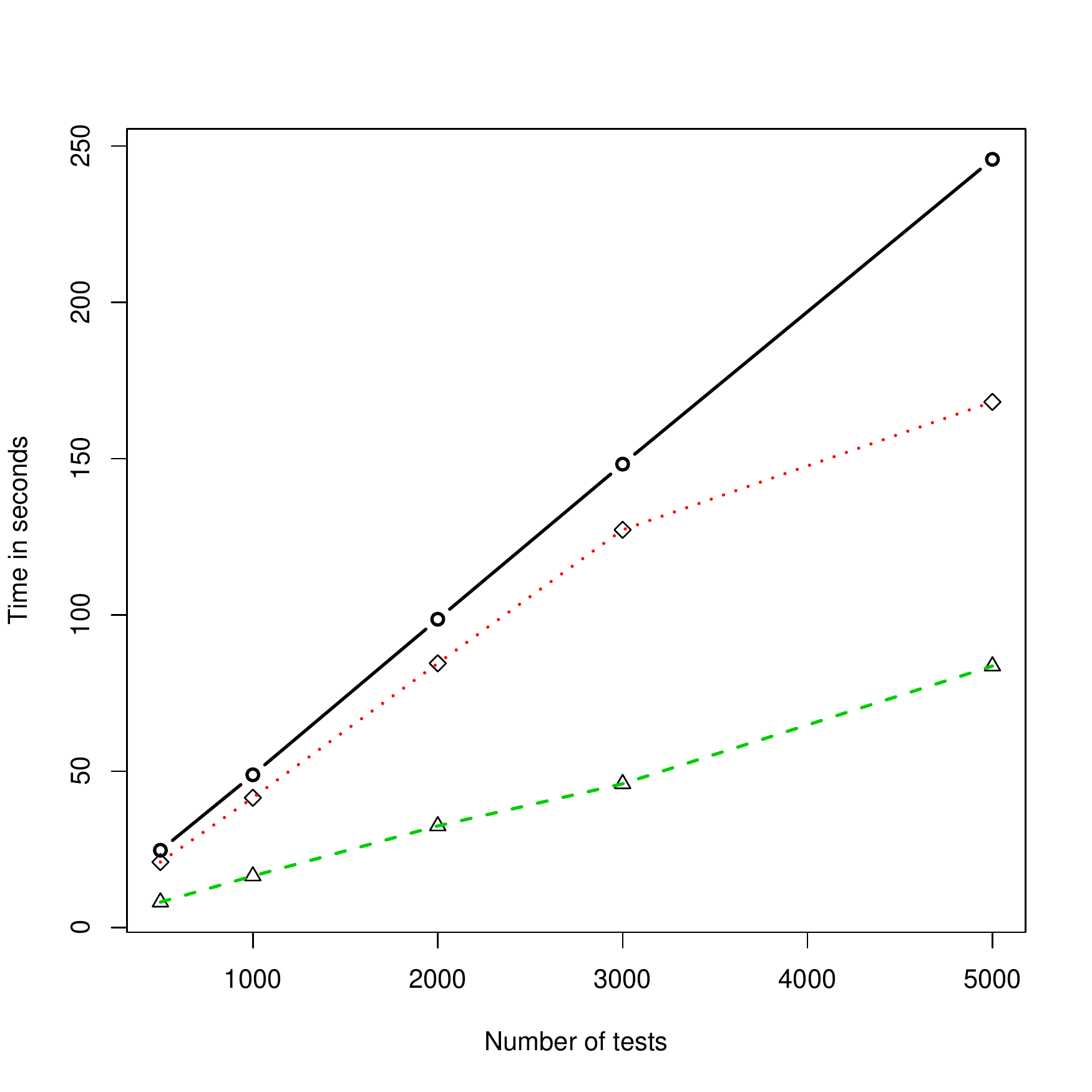}     &
\includegraphics[scale=0.25,trim=0 20 20 20]{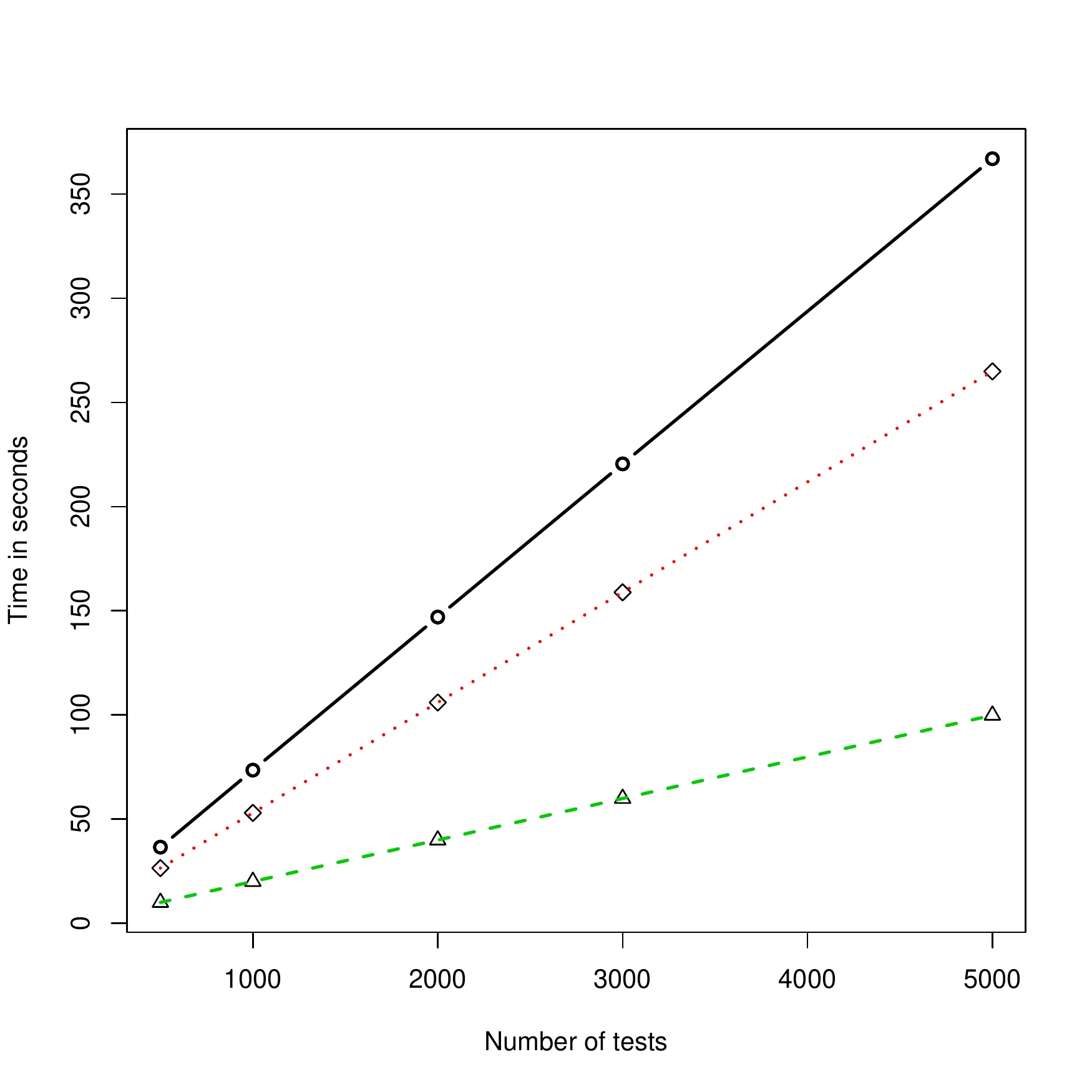}      &
\includegraphics[scale=0.25,trim=0 20 20 20]{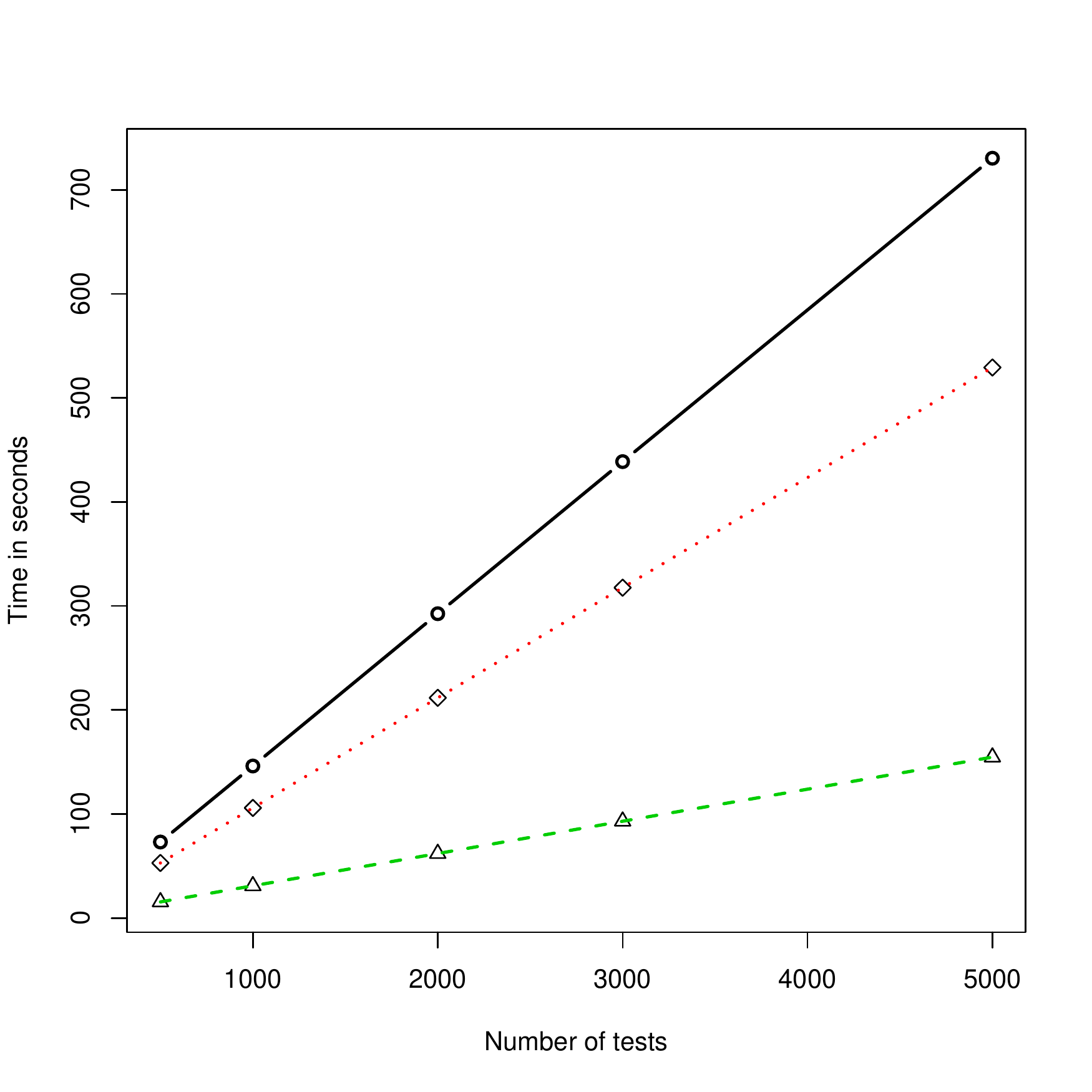}     \\
\footnotesize{$n=3000$}   &  \footnotesize{$n=5000$}  &  \footnotesize{$n=10000$}  \\  
\end{tabular}
\vspace{0.5cm}\parbox{12cm}{\small \centering Figure 1:\
Each point in all graphs corresponds to the average time (over 100 repetitions) for a given number of tests. The horizontal axis is the number of tests. The first row is the $G^2$ test conditioning on one variable, the second row conditioning on two variables and the third row conditioning on three variables. The green line with the triangle refers to the PLL, the black line with the circle refers to the gSquareDis and the red line with the diamond refers to the disCItest.}\vspace{-0.2cm}
\end{figure} 
\end{center}

When there is one conditioning variable, the disCItest is the fastest of all, for all three sample sizes. Note also, that PLL is the second fastest. When we increase the cardinality of the conditioning set to 2, we see that PLL changes position and becomes the fastest among them as the sample size increases. When there are three variables in the conditioning set, the PLL is clearly the fastest regardless of the sample size. In fact, for a give sample size (column-wise) the PLL has the smallest increase as the conditioning set increases. 

Tables 1, 2 and 3 contain the normalised time results, in which case the base time is that of PLL.  When the number for a given test is higher than 1 it means that the time required by that test is higher than the time required by PLL. The big differences appear in Table 3, where gSquareDis and disCItest are from 2 up to nearly 5 times slower than PLL. 

\begin{center}
\parbox{12cm}{\centering \small Table 1:\ Normalised times of gSquareDis and disCItest with one conditioning variable. The PLL has the value of 1.}
\small
\vspace{0.2cm} \tabcolsep=7.1truept
\begin{tabular}{c|cc|cc|cc}
\toprule
  & \multicolumn{6}{c}{Sample sizes} \\ \hline
        & \multicolumn{2}{c}{n=3000}  & \multicolumn{2}{c}{n=5000} & \multicolumn{2}{c}{n=10000}  \\
\midrule
Number of tests  & gSquareDis & disCItest  & gSquareDis & disCItest  & gSquareDis & disCItest \\[0.4mm] 
500  & 1.441 & 0.574 & 1.456 & 0.541 & 1.673 & 0.635 \\[0.4mm]
1000 & 1.476 & 0.590 & 1.476 & 0.557 & 1.667 & 0.632 \\[0.4mm]
2000 & 1.477 & 0.576 & 1.499 & 0.562 & 1.669 & 0.635 \\[0.4mm]
3000 & 1.429 & 0.568 & 1.487 & 0.557 & 1.666 & 0.637 \\[0.4mm]
5000 & 1.459 & 0.481 & 1.489 & 0.558 & 1.668 & 0.635 \\[0.4mm]
\bottomrule
\end{tabular}
\end{center}

\newpage
\begin{center}
\parbox{12cm}{\centering \small Table 2:\ Normalised times of gSquareDis and disCItest with two conditioning variables. The PLL has the value of 1.}
\small
\vspace{0.2cm} \tabcolsep=7.1truept
\begin{tabular}{c|cc|cc|cc}
\toprule
  & \multicolumn{6}{c}{Sample sizes} \\ \hline
        & \multicolumn{2}{c}{n=3000}  & \multicolumn{2}{c}{n=5000} & \multicolumn{2}{c}{n=10000}  \\
\midrule
Number of tests  & gSquareDis & disCItest  & gSquareDis & disCItest  & gSquareDis & disCItest \\[0.4mm] 
500  & 1.891 & 1.202 & 2.007 & 1.185 & 2.382 & 1.385 \\[0.4mm]
1000 & 1.875 & 1.158 & 2.067 & 1.181 & 2.360 & 1.376 \\[0.4mm]
2000 & 1.890 & 1.206 & 2.049 & 1.178 & 2.374 & 1.384 \\[0.4mm]
3000 & 1.888 & 1.186 & 2.050 & 1.178 & 2.376 & 1.387 \\[0.4mm]
5000 & 1.758 & 0.972 & 2.057 & 1.182 & 2.372 & 1.384 \\[0.4mm]
\bottomrule
\end{tabular}
\end{center}

\begin{center}
\parbox{12cm}{\centering \small Table 3:\ Normalised times of gSquareDis and disCItest with three conditioning variables. The PLL has the value of 1.}
\small
\vspace{0.2cm} \tabcolsep=7.1truept
\begin{tabular}{c|cc|cc|cc}
\toprule
  & \multicolumn{6}{c}{Sample sizes} \\ \hline
        & \multicolumn{2}{c}{n=3000}  & \multicolumn{2}{c}{n=5000} & \multicolumn{2}{c}{n=10000}  \\
\midrule
Number of tests  & gSquareDis & disCItest  & gSquareDis & disCItest  & gSquareDis & disCItest \\[0.4mm] 
500  & 3.050 & 2.583 & 2.007 & 2.647 & 4.712 & 3.412 \\[0.4mm] 
1000 & 2.968 & 2.522 & 2.067 & 2.643 & 4.720 & 3.418 \\[0.4mm] 
2000 & 3.036 & 2.601 & 2.049 & 2.652 & 4.718 & 3.414 \\[0.4mm] 
3000 & 3.225 & 2.768 & 2.050 & 2.652 & 4.714 & 3.412 \\[0.4mm] 
5000 & 2.939 & 2.010 & 2.057 & 2.653 & 4.722 & 3.422 \\[0.4mm] 
\bottomrule
\end{tabular}
\end{center}

\vspace{0.4cm} \noindent{\bf 4. Conclusions}\\
We have demonstrated, mathematically, how Poisson log-linear models can be used to test conditional independence. In addition, we have provided the relevant R function (see Appendix) which is based upon ready built-in functions in R. Both the $\chi^2$ and the $G^2$ tests are provided. The time comparisons have clearly favoured our function over the two functions available in the R package \textit{pcalg}. 

However, there is still room for improvement without moving to matlab, where the $G^2$ test is much faster. The PC algorithm (Spirtes et al., 2000) and the MMPC and MMHC algorithms Tsamardinos et al., 2006) are three classical examples where the $G^2$ test of independence is used. In all three algorithms the first step requires computation of all pairwise univariate associations. Performing the PLL in parallel, the first step only, will decrease the computational cost required by these algorithms for the network construction. 

The goal of the present manuscript was to point out that even with smart implementation and use of fast commands from other R packages and parallel computation, functions written in Java, Fortran or C++ will, obviously, be still faster than R. R may never reach these languages in terms of speed, yet functions can certainly be made to run faster. Indeed, the R package \href{https://cran.r-project.org/web/packages/Rfast/index.html}{Rfast} (Papadakis et al., 2017) is another example of speed. The code is written in C++ and the user calls it directly from R. This makes it extremely faster in comparison to R's commands. The time differences can be really extreme when many tests are to be performed, especially when many variables are present and all pairwise tests are required.

\vspace{0.4cm} \noindent{\bf Appendix: R function to calculate the $G^2$ and $\chi^2$ tests} \\
 
Below is the R function to perform the  $G^2$ and $\chi^2$ tests of (conditional) independence using PLL models. Note that the \textit{MASS} library is required for the command \textit{loglm}. For a univariate (uncoditional) association between two variables Pearson's chi squared test is performed.  

\begin{verbatim}
cat.ci <- function(xi, yi, cs, dataset) {
  ## the xi and yi are two numbers, 1 and 2 for example
  ## indicating the two variables whose conditional independence 
  ## will be tested
  ## xi, yi and cs must be different, non over-lapping numbers
  ## cs is one or more numbers indicating the conditioning variable(s)
  ## it is et to 0 by default. In this case an unconditional test of 
  ## independence is  performed
  ## dataset is the whole dataset, and is expected to be a matrix

  dataset = as.matrix(dataset)  ## makes sure it is a matrix 
  if ( sum(cs == 0) > 0 ) {  ## There are no conditioning variables
    ## a1 below contains the chi-square test, 
    a1 <- chisq.test(dataset[, xi], dataset[, yi], correct = FALSE) 
    ## faster than deriving it from PLL
    stat <- as.numeric( a1$statistic )
    dof <- as.numeric( a1$parameter )
    pval <- pchisq(stat, dof, lower.tail = FALSE, log.p = TRUE)
    res <- c( as.numeric(stat), pval, dof )  
    
  } else {   ## There are conditioning variables
    dat <- cbind( dataset[, c(xi, yi, cs)] ) 
    pa <- ncol(dat)
    colnames(dat) <- paste("V", 1:pa, sep = "")
    xnam <- paste("V", 3:pa, sep = "")
    form <- as.formula( paste("~ V1 + V2 ", paste(xnam, collapse = "+"), sep = "+") )
    mod <- xtabs(form , dat)  ## creates all the contingency tables 
    forma <- as.formula(paste( paste("~", "V1*", paste(xnam, collapse= "*"),
    sep = ""), paste("V2*", paste(xnam, collapse = "*"), sep = ""), sep = "+" ) )
    b1 <- summary( MASS::loglm(forma, mod) )$tests[1, 1:2]  ## PLL model
  }
  names(res) <- c("G-square test", "logged p-value", "df")
  res
}
\end{verbatim}

\vspace{0.4cm} \noindent{\bf   References}\\

\begin{description}

\item Agresti, A. (2002). \textit{Categorical data analysis}. John Wiley \& Sons.

\item Cheng, P. E., Liou, J.W., Liou, M., and Aston, J. A. (2006). Data information in contingency
tables: a fallacy of hierarchical loglinear models. \textit{Journal of Data Science}, \textbf{4}, 387–398.

\item Cheng, P. E., Liou, J.W., Liou, M., and Aston, J. A. (2007). Linear information models: An
introduction. \textit{Journal of Data Science}, \textbf{5}, 297–313.

\item Kalisch, M., Machler, M., Colombo, D., Maathuis, M. H., and Buhlmann, P. (2012). Causal
inference using graphical models with the r package pcalg. \textit{Journal of Statistical Software},
\textbf{47},1–26.

\item Neapolitan, R. E. et al. (2004). \textit{Learning Bayesian Networks}. Prentice Hall Upper Saddle
River.

\item Papadakis, M., Tsagris, M., Dimitriadis, M., Tsamardinos, I., Fasiolo, M., Borboudakis,
G., and Burkardt, J. (2017). Rfast: Fast R Functions. R package version 1.7.5.

\item Spirtes, P., Glymour, C. N., and Scheines, R. (2000). \textit{Causation, prediction, and search}. MIT
press.

\item Tsamardinos, I. and Borboudakis, G. (2010). Permutation testing improves bayesian network
learning. \textit{In Machine Learning and Knowledge Discovery in Databases}, 322–337. Springer.

\item Tsamardinos, I., Brown, L. E., and Aliferis, C. F. (2006). \textit{The max-min hill-climbing
bayesian network structure learning algorithm}. Machine learning, \textbf{65}, 31–78.

\item Zeileis, A., Wiel, M. A., Hornik, K., and Hothorn, T. (2008). Implementing a class of
permutation tests: the coin package. \textit{Journal of Statistical Software}, \textbf{28}, 1–23.

\end{description}

\footnotesize \centerline{Received December 2016; accepted March 2017.} \vspace{1cm}

\ni Michail Tsagris\\
mtsagris@yahoo.gr
Department of Computer Science\\
University of Crete\\
Herakleion 71305, Greece\\

\end{document}